\documentclass[12pt]{article}%
\usepackage{amsfonts}
\usepackage{amsmath}
\usepackage{amssymb}
\usepackage{graphicx}%
\providecommand{\U}[1]{\protect\rule{.1in}{.1in}}

\begin{document}

\title{Gravitational waves at their own gravitational speed}
\author{C. S. Unnikrishnan$^{1}$ and George T. Gillies$^{2}$\\$^{1}$\textit{Gravitation Group, Tata Institute of Fundamental Research, }\\\textit{\ Homi Bhabha Road, Mumbai - 400 005, India}\\$^{2}$\textit{School of Engineering and Applied Science,}\\\textit{University of Virginia}, \textit{Charlottesville, VA 22904-4746, USA}\\E-mail address: $^{1}$unni@tifr.res.in, $^{2}$gtg@virginia.edu\\\medskip\\This essay received honorable mention in the `Gravity Research\\Foundation 2018 Awards for Essays on Gravitation'. }
\date{}
\maketitle

\begin{abstract}
Gravitational waves propagate at the speed of light in general relativity,
because of its special relativistic basis. However, light propagation is
linked to the electromagnetic phenomena, with the permittivity and
permeability constants as the determining factors. Is there a deeper reason
why waves in a geometric theory of gravity propagate at a speed determined by
electromagnetic constants? What is the relation between gravity's own
constants and the speed of gravitational waves? Our attempt to answer these
fundamental questions takes us far and deep into the universe.

\pagebreak

\end{abstract}

The first guesses that light was linked to the phenomenon of electricity and
magnetism were after Wilhelm Weber's formulation of electrodynamics in the
1850s that unified the experimental inputs from Coulomb and Amp\`{e}re
\cite{weber,assis}. Maxwell's theoretical triumph in finding the unified
equations and their wave solution established that the speed of light was
determined by the coupling constants of electricity and magnetism,
$c^{2}=1/\varepsilon_{0}\mu_{0}.$ Therefore, \textit{the speed of light could
be determined purely from measurements of forces in electrostatics and
magnetostatics }\cite{maxwell}. We want to pursue the similar natural question
of whether or not we can determine the nature of propagation of gravitational
waves \textit{entirely from the measured forces of gravity and
gravitomagnetism, or from the forces of geometro-statics.}

This is important conceptually and physically because there is no reason why
waves in the physical or geometric theory of gravity propagate at a speed
determined by the constants of another interaction, electromagnetism.
Gravitational waves have some similarities and even more dissimilarities to
electromagnetic waves. After all, one should have been able to build the
fundamental theory of gravity without referring to the `speed of light'. The
historical priority to the electromagnetic reference is through the common
relativistic basis. What is the relation between gravity's own constants and
the speed of gravitational waves? Being the waves described in a geometric
theory of gravity, with the Newtonian coupling constant $G$, their propagation
should be determined entirely by gravitational coupling and, perhaps, some
geometrical considerations. Conventionally, the theoretical reason is taken
for granted, and quoted often, that because of the special relativistic basis
of general relativity, the propagation of weak gravitational waves in
linearized general relativity is at the speed of light. The physical surprise
remains why the propagation of the waves of gravity are determined by
electromagnetic constants. However, had we already known of the existence and
propagation of gravitational waves at a time when the link between
electromagnetism and light was established, before 1900, then surely we would
have searched for a physical link between electromagnetism and gravity, and
also a relation between `$c$' and the gravitational coupling.

The electric permittivity of empty space, of the old ether, can be determined
from the measurement of the magnitude of the force between known charges,
using the Coulomb's law,%
\begin{equation}
F=k_{e}\frac{Q_{1}Q_{2}}{r^{2}}=\frac{1}{4\pi\varepsilon_{0}}\frac{Q_{1}Q_{2}%
}{r^{2}}%
\end{equation}

Similarly, the force between two `magnetic poles' is
\begin{equation}
F=k_{m}\frac{p_{1}p_{2}}{r^{2}}=\frac{\mu_{0}}{4\pi}\frac{p_{1}p_{2}}{r^{2}}%
\end{equation}
The magnetic permeability constant $\mu_{0}$ is measured from the force
between current carrying conductors separated by $d$, using Amp\`{e}re's law
for the force between two currents,
\begin{equation}
F=-\frac{\mu_{0}I_{1}I_{2}}{2\pi d}%
\end{equation}

After the realization that the ratio of the Coulomb constant and the
Amp\`{e}re constant -- the ratio of the electric and magnetic force constants
-- $k_{e}/(\mu_{0}/4\pi)=1/\varepsilon_{0}\mu_{0}$ is equal to the square of
the velocity of light $c^{2},$ modern measurements are defined from just the
value of `$c$', with $\mu_{0}$ defined as $4\pi\times10^{-7}~kgm/C^{2}.$ What
are the relevant similar constants of gravity that determine the speed of
gravitational waves and how do we measure them?

Well before the study of the electric force itself, the `electric' aspect of
gravity was studied extensively after the discovery of Newtonian gravitation,
culminating in the measurement of the constant $G$ by Cavendish from the
famous inverse square law
\begin{equation}
F=-G\frac{m_{1}m_{2}}{r^{2}}%
\end{equation}
But there was no `magnetic gravity' seen in nature and nobody guessed or
expected it, despite some theoretical clues, till general relativity predicted
it. Gravitomagnetism, since then, has been well studied
\cite{Einstein,deSitter,Schiff,Thorne,Cuifolini}. But direct measurements
became possible only recently \cite{Lageos,gpB}.

We now explore the possibility of expressing the velocity of the gravitational
waves, confirmed to be the numerical value `$c$' from the joint LIGO-Fermi
observation of the gravitational waves and light from the binary neutron stars
merger \cite{LIGO-bns}, \textit{entirely in terms of the force constants from
gravitational measurements of the `electric' and `magnetic' effects in
gravity}. This will free the physical existence of waves of the gravitational
fields from the link to electrodynamic phenomena as well as the theoretical
structure of relativity, Lorentz transformations etc. While all interactions
are subject to dynamics allowed by relativity, each should be describable
entirely in terms of its intrinsic constants, much like the situation with
electromagnetic waves. The equivalent of permittivity in gravity is of course
$1/G,$ measured in the Cavendish experiment, which was similar to the Coulomb
experiment. We need a measurement of gravitomagnetism between known
gravitomagnetic charges to get a value of the gravitational permeability
$\mu_{G}$. Ideally, this would be an Amp\`{e}re type experiment in which
forces between currents of the charges of gravity, or mass currents, are
measured. However, in the absence of that possibility, the next ideal
measurement is of the gravitomagnetic interaction between spinning objects
because spin, or angular momentum, is the closed current of the charge of
gravity This is available from the direct measurement of the gravitomagnetic
precession of the gyroscope in the Gravity Probe-B experiment \cite{gpB}, and
from the analysis of the precession of orbits of the LAGEOS satellites
\cite{Lageos}. We use the direct measurement of the gyroscope precession in
the GP-B experiment. \ While these measurements have relatively large
systematic errors, they are nevertheless the only direct measurements
available and are quite suitable for our purpose. Since we know the precession
rate and the `gravitomagnetic moment' (i.e., the angular momentum) of the
source, which is the Earth, we can estimate $\mu_{G}.$ \ The gravitomagnetic
interaction between the gyroscope with spin $s$ and the earth with angular
momentum $J$ is
\begin{equation}
E=\mu_{G}s\cdot J/R^{3}%
\end{equation}
The precession of the gyroscope, averaged over an orbit assuming a circular
polar orbit \cite{gpB} is
\begin{equation}
\left\langle \omega\right\rangle =\frac{\mu_{G}J}{2R^{3}}=\frac{\mu_{G}%
I\Omega}{2R^{3}}\simeq\frac{\mu_{G}(0.33)MR_{e}^{2}\Omega}{2R^{3}}%
\end{equation}
The normalized moment of inertia $I/MR^{2}=0.33$ is from the earth model
\cite{Jeffreys} and $J\simeq6\times10^{33}kgm^{2}/s.$

The Gravity Probe B orbit was at $R=R_{E}+640~km.$ From the measured (average
over 4 gyros) precession rate, $37.2\pm7.2$ milli-arcseconds/year,
$(5.7\pm1.1)\times10^{-15}$ radian/s, we get
\begin{equation}
\mu_{G}\simeq6.7\left(  \pm1.3\right)  \times10^{-28}~m/kg
\end{equation}
We find, $\mu_{G}\varepsilon_{G}=\mu_{G}/G\simeq(1.0\pm0.2)\times
10^{-17}~s^{2}/m^{2}.$ $\ $Thus, $\left(  \mu_{G}/G\right)  ^{-1}\simeq\left(
9.9\pm1.9\right)  \times10^{16}~m^{2}/s^{2}\simeq c_{g}^{2}.$

This is a very satisfactory result. We have a description of the relativistic
gravitational effects in the wave sector in terms of the purely gravitational
constants $G$ and $\mu_{G},$ much like the description of electromagnetic
effects in terms of $\varepsilon_{0}$ and $\mu_{0}.$ We did not use the
constant `$c$' in our analysis, but verified that the product $G/\mu_{G}\simeq
c_{g}^{2},$ the square of a velocity close to the empirically known velocity
of gravitational waves. That it is also the velocity of light, of
electromagnetism, reflects the universality of relativity. This exercise was
similar to W. Weber's analysis in which he found that the product of the
electric and magnetic force constants were related to the known velocity of light.

We have deeper physics linking gravity and electromagnetism, to be explored in
the new relation,
\begin{equation}
G/\mu_{G}=1/\varepsilon_{0}\mu_{0}%
\end{equation}
Since the values of `$c$' and `$\mu_{0}$' are exactly defined in the modern
system, and with one equation, there is only one independent fundamental
constant -- $G$ -- for the wave sector of the two long range interactions,
gravity and electromagnetism. This shows the need and importance of
measurements of $G$ at higher precision.

The inverse of the gravitational permeability with units $kg/m$ and its
enormous value is suggestive of a `massive compact' region. Though it is a
familiar relation in the context of black holes, the correct identification,
however, leads us to a surprise. Since $\mu_{G}$ is fundamental constant, it
should be associated with the universe itself, and we have the relation
\begin{equation}
1/\mu_{G}\simeq2M_{u}/R\simeq2\frac{4\pi}{3}\rho_{u}R^{2}=\frac{8\pi}{3}%
\rho_{u}R^{2}%
\end{equation}
where $\rho_{u}$ is the average density of the universe and $R$ is a length
scale of the order of the Hubble radius. Thus the product of gravity's
constants, $G$ and $1/\mu_{G}$ is like the gravitational potential of the
matter-energy of the entire universe, and it is numerically equal to
$c_{g}^{2}$. This indicates the decisive role of the matter-energy content of
the universe in relativistic physics \cite{cosrel}.

We conclude that there is a self-contained description of the propagation of
gravitational waves in terms of only the constants of gravity, without linking
them to the electromagnetic constants and the speed of light. Gravitational
waves propagate at a speed determined by its interaction constants, $G$ and
$\mu_{G}$, entirely independently measurable in the laboratory and in
space-based gravitation experiments. The two measurements we needed were the
Cavendish measurement of $G,$ the analog of $1/\varepsilon_{0},$ and the
direct measurement of the gravitomagnetic interaction between the spinning
earth and a gyroscope, which gave us the gravitomagnetic permeability $\mu
_{G}.$ The link $\mu_{G}/G=\mu_{0}\varepsilon_{0}$ is through the nature's
implementation of the principle of relativity. As not entirely unexpected, the
matter-energy content in the universe plays a decisive role in gravitomagnetic
effects, with a Machian flavor, and invites us for further exploration.

\end{document}